\documentclass[10pt]{iopart}

\DeclareMathAlphabet\mathbfcal{OMS}{cmsy}{b}{n}
\usepackage{mathrsfs}
\usepackage{bm,setstack,amssymb}
\usepackage{iopams}
\usepackage{epsfig}
\usepackage{amssymb}
\usepackage{amsthm}
\usepackage{graphicx}
\usepackage{color}
\usepackage{cite}
\usepackage[dvipsnames]{xcolor}
\usepackage{adjustbox}
\usepackage{dcolumn}
\usepackage{subfigure}
\usepackage{color}
\usepackage[colorlinks=true,linkcolor=blue]{hyperref}

\def\vecb#1{\boldsymbol{#1}}

\mathchardef\mhyphen="2D

\def\ket#1{|#1\rangle}

\def\scal#1#2{\langle#1|#2\rangle}
\def\matr#1#2#3{\langle#1|#2|#3\rangle}
\def\eqref#1{\eref{#1}}
\def\uvo#1{\lq\lq #1\rq\rq}
\def\ii{{\rm i}}
\def\ee{{\rm e}}

\begin{document}
	
	\title{Analytic approach to the Landau-Zener problem in bounded parameter space}
	\author{Felipe Matus\footnote{Author to whom any correspondence should be addressed.}, Jan St{\v r}ele{\v c}ek, Pavel Cejnar}
	\address{Institute of Particle and Nuclear Physics, Faculty of Mathematics and Physics, Charles University,
		V Hole{\v s}ovi{\v c}k{\' a}ch 2, 180 00 Prague, Czechia}
	\ead{\mailto{matus@ipnp.mff.cuni.cz}, \mailto{strelecek@ipnp.mff.cuni.cz}, \mailto{cejnar@ipnp.mff.cuni.cz}}
	
	\date{\today}
	
	\begin{abstract}
	Three analytic solutions to the Schr{\" o}dinger equation for the time-dependent Landau-Zener Hamiltonian are presented. They correspond to specific finite-time driving paths in a bounded parameter space of a two-level system. Two of these paths go through the avoided crossing of levels, either with a constant speed or with variable speed that decreases in the region of reduced energy gap, the third path bypasses the crossing such that the energy gap remains constant. The solutions yield exact time dependencies of the excitation probability for the system evolving from the ground state of the initial Hamiltonian. The Landau-Zener formula emerges as an approximation valid within a certain interval of driving times for the constant-speed driving through the avoided crossing. For long driving times, all solutions converge to the prediction of the adiabatic perturbation theory. The excitation probability vanishes at some discrete time instants. 
	\end{abstract}
	
	
	\maketitle
	
	\section{Introduction}\label{sec:intro}
	
	The evolution of quantum systems driven by time-dependent Hamiltonians lies in foundations of diverse physical processes. If the system starts in one of the non-degenerate eigenstates of the initial Hamiltonian and if the subsequent Hamiltonian variation is very slow, the adiabatic theorem of quantum mechanics~\cite{Born28} ensures that the system remains in the respective eigenstate of the evolving Hamiltonian. Such adiabatic evolution is accompanied by the appearance of dynamic and geometric phase factors~\cite{Berr84}. However, if the Hamiltonian change is not sufficiently slow, transitions between instantaneous eigenstates are unavoidable. Since variations of individual eigenvectors proceed in the fastest way for the states separated by small energy gaps, such transitions arise most likely when the energy levels, which carry non-negligible immediate population probabilities, happen to be energetically close to other, less populated levels. The most common approach to simulate such pairwise nonadiabatic effects makes use of a~two-level approximation. The celebrated Landau-Zener (LZ) model (also independently introduced by St{\" u}ckelberg and Majorana) \cite{Land32,Zene32,Stue32,Majo32} describes a~two-state quantum system driven by an evolving two-by-two Hamiltonian through an avoided crossing of levels in an infinite parameter and time domain, and provides a~valid estimate of the final population of both states.
	
	The LZ approximation is employed in the description of a~vast variety of systems and phenomena. To give just some examples of applications and a sample of relevant references, we mention the use of the LZ approach in atomic and molecular collisions \cite{Lee79,Ostro91,Niki99}, nuclear fission and other nuclear reactions \cite{Hill53,Cind86,Thie90,Mire07}, quantum dissipative processes \cite{Wilk88,Ao91,Wubs06,Sait07}, effects related to quantum dots and Bose-Einstein condensation \cite{Sini02,Zwanenburg,Cao,Sun08,Ribe09,Urda13,Mall22}, cold and ultracold molecules \cite{Kohler_cold,Mark,Lang}. The LZ model provides an adequate treatment of dynamical aspects of cosmological and condensed-matter quantum phase transitions in connection with the Kibble-Zurek mechanism \cite{Kibb80,Zure85,Dams05,Dzia05}. Understanding of nonadiabatic effects through the LZ mechanism is essential for recent attempts to control the dynamics of quantum systems in different areas of science, like in the coherent manipulation of molecular systems~\cite{Rice00}, quantum computation~\cite{Alba18}, quantum information processing~\cite{Luce08,Izmalkov,Dupont, Pla13} and many others. Therefore, various analytic or approximate solutions related to the original and generalized LZ problem are of great interest. Various kinds of such solutions were discussed in the literature, see, e.g., references~\cite{Carr85,Rosen,Nikitin1,Demkov,Wann65,Kaya84,Mull89,Vitanov2,Vita99,Yan10} for two-level systems and \cite{Carr86,Demk01,Pokr02,Volk04,Vasi07,Sini13,Sini15} for generalizations to multilevel systems. A very recent detailed review of various applications of the LZ problem can be found in \cite{Oleh}.
	
	In this paper, we derive analytic solutions to the LZ dynamics in finite time for three specific driving paths in bounded parameter space of the two-state model. While one path is based on the linear LZ drive, the other two paths are based on the geometry of the ground state \cite{Tom06,Reza10}, which can be used to study for example quantum phase transitions \cite{Zana06,Buon07} and became an appealing object of interest in quantum information theory \cite{Zana07}. The idea of geometry inspires new driving paths, which bound from below the minimum driving time to reach maximum fidelity for sufficiently slow protocols\cite{Buko19}. Moreover, geometry-inspired drivings appear naturally when considering local adiabatic-evolution approaches \cite{Rola02}. The three solutions introduced in this paper will be used to extend our recent numerical analysis of quantum driving in two- and multi-state qubit systems in \cite{We}, where it was demonstrated the existence of a critical driving time at which the system makes a sharp crossover from the regime described by the LZ approximation to the asymptotic-time regime governed by the adiabatic perturbation theory (APT) of Rigolin, Ortiz and Ponce~\cite{Ortiz}. Our present analytic solutions provide exact description of the dynamics on both sides of this crossover and also in the cases when the LZ approximation cannot be used at all. These results can be experimentally tested by realizing an effective two-level system using ultracold atoms forming a Bose-Einstein condensate in an accelerated optical lattice, which allows stable control of the system parameters \cite{Zene09,Baso11}.
	
	The paper is structured as follows: In section~\ref{sec:2} we introduce the LZ Hamiltonian and describe the three relevant driving paths. In section~\ref{sec:3} we derive analytic solutions to the time-dependent Schr{\" o}dinger equation for these paths. In section~\ref{sec:4} we extract from these solutions the evolution of the excitation probability (infidelity). In section~\ref{sec:5} we discuss the analytic expression of the final infidelity and compare it with predictions of the APT. In section~\ref{sec:6} we give a brief summary. To keep the paper self-contained, we attach three appendices encapsulating some technical prerequisites.

\section{The Hamiltonian and driving paths}
\label{sec:2}

We analyze a coherent driving in the parameter space of a two-level Hamiltonian
\begin{equation}\label{HLZ}
	\hspace{-1cm}
	\hat{H}_{\rm LZ}(\boldsymbol{r}) =  -\boldsymbol{r}\cdot \hat{\boldsymbol{\sigma}} = 
	-\left(\matrix{%
		z & x-\ii y\cr
		x+\ii y & -z \cr}\right)
	= -r
	\left(\matrix{%
		\cos\vartheta & \ee^{-\ii\varphi}\sin\vartheta\cr
		\ee^{\ii\varphi}\sin\vartheta& -\cos\vartheta \cr}\right),
\end{equation}
where $\boldsymbol{\sigma}=(\hat{\sigma}_x,\hat{\sigma}_y,\hat{\sigma}_z)$ are the three Pauli matrices and $\boldsymbol{r}= (x,y,z)=r(\sin\vartheta\cos\varphi,\sin\vartheta\sin\varphi,\cos\vartheta)$ denote three real control parameters of the model written in the form of Cartesian and spherical coordinates, respectively. 
We have $x,y,z\in(-\infty,+\infty)$ and $r\in[0,\infty)$, $\vartheta\in[0,\pi]$, $\varphi\in[0,2\pi)$.
Formula \eqref{HLZ} represents the most general Hamiltonian in dimension 2 up to a possible overall shift of the spectrum.
It can describe, e.g., a spin-$\frac{1}{2}$ particle in a magnetic field $\boldsymbol{B}\propto\boldsymbol{r}$. 
We will further refer to \eqref{HLZ} as to the LZ Hamiltonian.
We note that this Hamiltonian and its parameters are considered dimensionless. The Hamiltonian of any particular implementation of the two-level system reads as ${\epsilon\hat{H}_{\rm LZ}(\boldsymbol{r})=\hat{H}_{\rm LZ}(\epsilon\boldsymbol{r})}$, where the constant $\epsilon$ sets the relevant energy scale of the actual problem. Therefore, the time variable in the following calculations (sections \ref{sec:3}--\ref{sec:5}) is always expressed in units of $\hbar/\epsilon$.

The eigenvalues and the corresponding normalized eigenvectors of the LZ Hamiltonian read as
\begin{eqnarray}
	E_0(\boldsymbol{r})=-r,\quad\ket{E_0(\boldsymbol{r})} &= 
	\left(\matrix{%
		\cos\frac{\vartheta}{2} \cr
		\ee^{\ii\varphi} \sin\frac{\vartheta}{2}\cr}\right),
	\label{gs2x2}\\
	E_1(\boldsymbol{r})=+r,\quad\ket{E_1(\boldsymbol{r})} &= 
	\left(\matrix{%
		-\ee^{-\ii\varphi}\sin\frac{\vartheta}{2} \cr
		\cos\frac{\vartheta}{2}\cr}\right),
	\label{es2x2}
\end{eqnarray}
where $E_0$ represents the ground state and $E_1$ the excited state.
Since $\Delta E(\boldsymbol{r})={E_1(\boldsymbol{r})-E_0(\boldsymbol{r})}=2r$, the system becomes degenerate in the diabolic point at ${r=0}$.
Because $r$ is just a scaling parameter of Hamiltonian \eqref{HLZ}, the eigenvectors \eqref{gs2x2} and \eqref{es2x2} depend only on angles~$\vartheta$ and $\varphi$. 
The eigenvectors $\ket{E_0(\boldsymbol{r})}$ and $\ket{E_1(\boldsymbol{r})}$ for any $\vartheta,\varphi$ can be obtained from those for ${\vartheta=0}$ ($\varphi$ arbitrary) by a  unitary transformation:
\begin{equation}
	\ket{E_n(\boldsymbol{r})}=\ee^{\ii\left(\sin\varphi\,\hat{\sigma}_x-\cos\varphi\,\hat{\sigma}_y\right)\vartheta/2}\ket{E_n(\vartheta=0)},\quad n=0,1.
	\label{simil}
\end{equation}
This allows one to explicitly construct the unitary transformation connecting any pair of Hamiltonians $\hat{H}_{\rm LZ}(\boldsymbol{r})$ and $\hat{H}_{\rm LZ}(\boldsymbol{r}')$ satisfying $r=r'$.

\begin{figure}[t!]
\centering
\includegraphics[width=\linewidth]{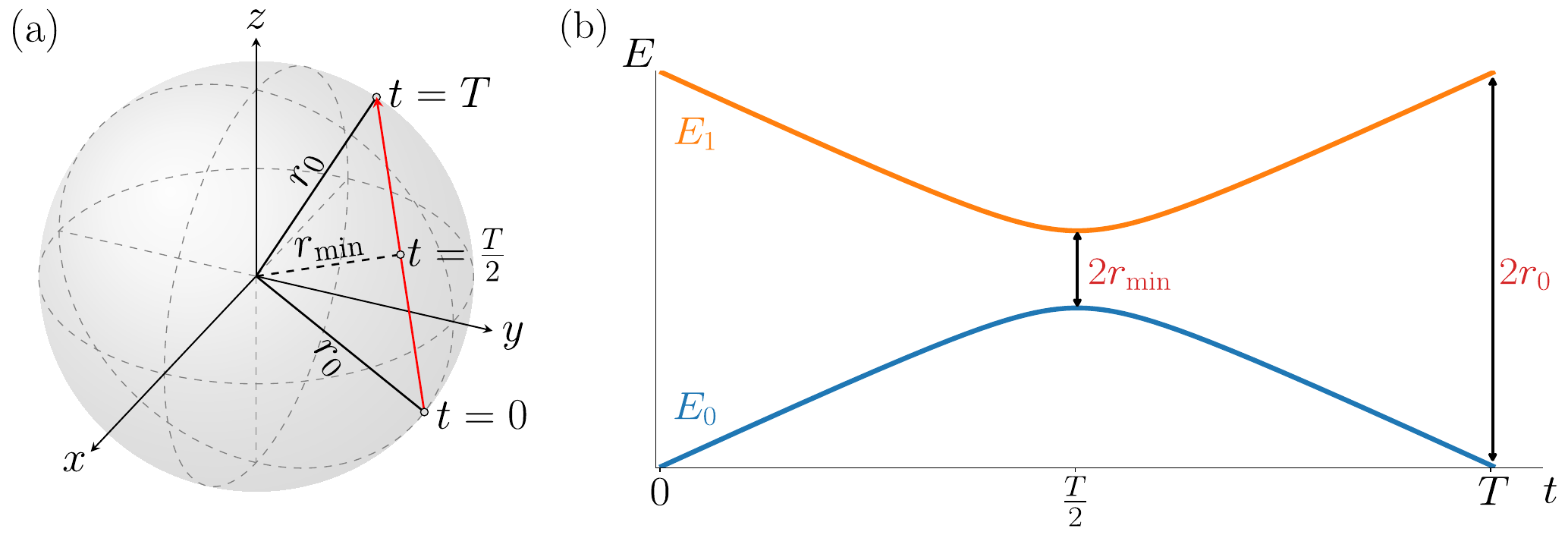}
\caption{(a) Restricted parameter space of the LZ Hamiltonian \eqref{HLZ} formed by a ball of radius $r_0$, with a driving trajectory along a chord of the sphere. (b) The instantaneous spectrum of the LZ Hamiltonian corresponding to a constant-speed driving along the chord trajectory.}
\label{fig:Figure1}
\end{figure}

The original formulation of the Landau-Zener problem \cite{Land32,Zene32,Stue32,Majo32} considers an external driving of the two level system along an infinite line $\boldsymbol{r}(t)=\boldsymbol{r}(0)+\boldsymbol{u}t$ in the parameter space, where $t$ is time and $\boldsymbol{u}$ is a finite driving velocity. The system is initiated in infinite past $t\to-\infty$ in the ground state of the Hamiltonian at $\boldsymbol{r}(-\infty)$ and then evolved with $t\in(-\infty,+\infty)$ by the time-dependent Hamiltonian $\hat{H}_{\rm LZ}(\boldsymbol{r}(t))$, passing the region of avoided crossing of both levels. The energy gap is minimal at $t=0$, when the line gets to the smallest distance ${r_{\min}=|\boldsymbol{r}(0)|}$ to the diabolic point at ${r=0}$. The task is to determine the probability of finding the system in its ground state at $\boldsymbol{r}(+\infty)$ in infinite future $t\to+\infty$. Note that due to the independence of eigenvectors \eqref{gs2x2} and \eqref{es2x2} on the size of $\boldsymbol{r}$, the initial and final ground states can be written as $\ket{E_0(-\boldsymbol{u}t_0)}$ and $\ket{E_0(+\boldsymbol{u}t_0)}$, respectively, where $t_0$ is an arbitrary time constant.

In the present paper, we consider a finite-time driving in a \emph{bounded} parameter domain. We assume that the initial parameter value $\boldsymbol{r}_{\rm I}={(x_{\rm I},y_{\rm I},z_{\rm I})}$ is set at time ${t_{\rm I}=0}$ and the final value $\boldsymbol{r}_{\rm F}={(x_{\rm F},y_{\rm F},z_{\rm F})}$ is reached at time ${t_{\rm F}=T}$. So $T$ is the overall duration of the drive and $t\in[0,T]$. A sketch of driving along a line segment embedded in the parameter sphere of radius $r_0$ is shown in figure~\ref{fig:Figure1}. Panel (a) depicts the restricted driving path---a line segment forming a chord of the sphere---and panel (b) plots the corresponding evolution of the two-level spectrum. Without loss of generality we choose ${x_{\rm I}=x_{\rm F}}\equiv{x_0>0}$, ${y_{\rm I}=y_{\rm F}=0}$ and ${-z_{\rm I}=z_{\rm F}}\equiv{z_0>0}$. 

\begin{figure}[t!]
\centering
	\begin{adjustbox}{right}
	\includegraphics[width=0.6\linewidth]{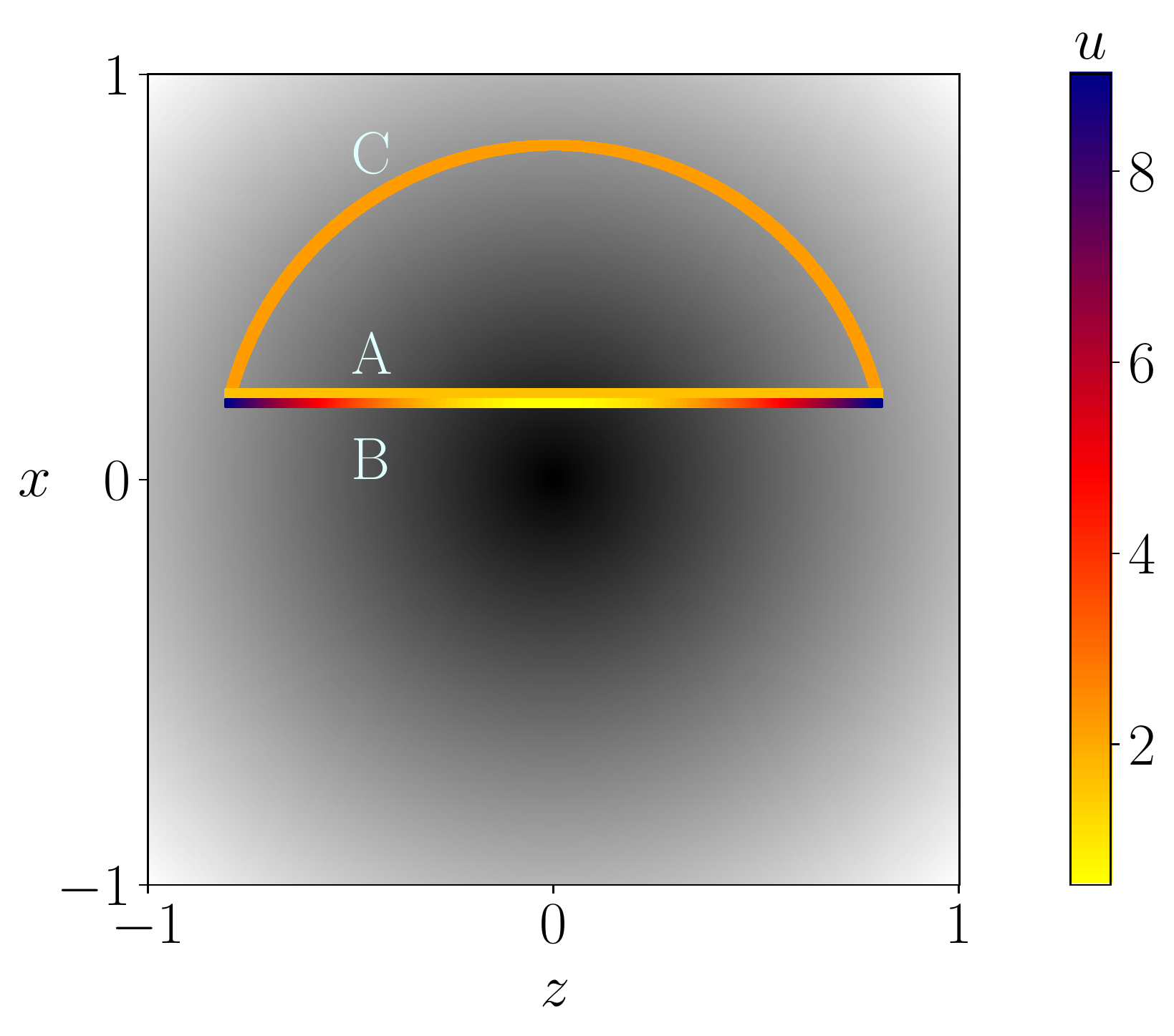}
	\end{adjustbox}
\caption{The three types of paths in the parameter space of Hamiltonian \eqref{HLZ} studied in this work. The plain speed from \eqref{speed_plane} evaluated for the total driving time ${T=1}$ is expressed by the color scale indicated on the right, the energy gap $\Delta E$ by the shades of gray of the background (darker shades correspond to smaller gaps). Constants $x_0$ and $z_0$ are chosen arbitrarily, $y$ is set to 0.}
	\label{fig:Figure2}
\end{figure}

We consider three specific drivings between the points $\boldsymbol{r}_{\rm I}$ and $\boldsymbol{r}_{\rm F}$. These drivings were studied numerically in our previous work \cite{We}.  We use the term \lq\lq path\rq\rq\ to specify the full time dependence $\boldsymbol{r}(t)$ for $t\in[0,T]$. While the first path is a straightforward conversion of the $\boldsymbol{u}={\rm const.}$ driving of the LZ type to the restricted parameter space, the other two paths represent generalizations of the LZ driving to the forms with $\boldsymbol{u}\neq{\rm const}$. As will be explained later, the reason for choosing the latter two paths lies in the fact that they are somehow privileged in terms of the geometry of the ground state \cite{Tom06,Reza10,Zana06,Buon07,Zana07,Buko19}. The three paths are defined as follows:
\begin{itemize}	
	\item
	Path $\rm{A}$: a linear change of parameter $z$ along the line with ${y=0}$ and ${x=x_0}$, i.e.,
	\begin{equation}\label{pathA}
		x(t)=x_0,\ y(t)=0,\ z(t)=z_0\left(\frac{2t}{T}\!-\!1\right).
	\end{equation}
	So the speed of driving along this line is constant, equal to ${u=2z_0/T}$. 	
	\item
	Path $\rm{B}$: driving along the same line as in path~$\rm{A}$, but with a variable speed $u(t)$, 
	\begin{eqnarray}\label{pathB}
		x(t)=x_0,\ y(t)=0,\ z(t)=x_0\tan \alpha(t),
		\\
		\label{angleBC}
		\alpha(t) = \alpha_0 \left(\frac{2t}{T}-1\right), \quad \alpha_0=\arctan\left(\frac{z_0}{x_0}\right).
	\end{eqnarray}
	The speed decreases in the small-gap region as $u(t)=\alpha_0\, \Delta E(\boldsymbol{r}(t))^2/(2x_0T)$.	
	\item
	Path $\rm{C}$: driving with a constant speed $u=2\alpha_0r_0/T$ in the plane ${z\times x}$ along the upper circular segment connecting the initial and final points, so
	\begin{equation}\label{pathC}
		\quad\quad\!  x(t)=r_0 \cos\alpha(t),\ y(t)=0,\ z(t)=r_0 \sin\alpha(t),
	\end{equation} 
	where $\alpha(t)$ is defined in \eqref{angleBC} and $r_0 = \sqrt{x_0^2 + z_0^2}$.	
\end{itemize}
These paths are drawn in figure~\ref{fig:Figure2}. We note that while paths~$\rm{A}$ and~$\rm{B}$ proceed through the avoided-level crossing shown on figure~\ref{fig:Figure1}, path~$\rm{C}$ bypasses the avoided crossing region so that it maintains the energy gap $\Delta E$ constant.
It needs to be stressed that the results obtained below would be the same if the paths are rotated by arbitrary angles around the origin ${r=0}$ in the space ${x\times y\times z}$. 

The immediate size of driving speed in the parameter space is given by
\begin{equation}\label{speed_plane}
	u(t)=\sqrt{\dot{x}(t)^2+\dot{y}(t)^2+\dot{z}(t)^2},
\end{equation} 
where dots denote time derivatives.
An alternative definition of speed follows from the Provost-Vallee metric that can be introduced for any parameter-dependent quantum system with discrete energy levels \cite{Provost}.
In our two-level system, the Provost-Vallee metric tensor reads as
\begin{equation}\label{metric}
	g_{\mu\nu}(\boldsymbol{r})={\rm Re}\frac
	{\matr{E_{0}(\boldsymbol{r})}{\frac{\partial\hat{H}_{\rm LZ}}{\partial\mu}(\boldsymbol{r})}{E_{1}(\boldsymbol{r})}\matr{E_{1}(\boldsymbol{r})}{\frac{\partial\hat{H}_{\rm LZ}}{\partial\nu}(\boldsymbol{r})}{E_{0}(\boldsymbol{r})}}
	{\bigl(E_1(\boldsymbol{r})-E_0(\boldsymbol{r})\bigr)^2},
\end{equation}
where $\mu,\nu\in\{x,y,z\}$ or $\{r,\vartheta,\varphi\}$.
Let us note that for a general Hamiltonian the metric tensor can be assigned separately to each parameter-dependent eigenstate $\ket{E_n}$, but in the two-level systems the tensors of both levels coincide.
This approach invokes the language of Riemannian geometry of curved spaces \cite{Pokolnikov2}. 
In particular, the tensor $g_{\mu\nu}$ induces a metric on the parameter space, which introduces the notion of length $\ell=\int \sqrt{g_{\mu\nu}d\mu d\nu}$ of any driving trajectory and the respective speed
\begin{equation}\label{speed_on_manifold}
	v(t)=\sqrt{\sum_{\mu,\nu}g_{\mu\nu}\bigl(\boldsymbol{r}(t)\bigr)\,\dot{\mu}(t)\,\dot{\nu}(t)}.
\end{equation}
In this article, these two quantities described with respect to the metric space of parameter will be referred to as the metric length and metric speed, respectively. 
Now we are ready to explain that the plain speed~\eqref{speed_plane} for path~$\rm{B}$ in \eqref{pathB} is varied so that the metric speed~\eqref{speed_on_manifold} remains constant, equal to ${v = \alpha_0/T}$.
Hence, in accord with \eqref{metric}, the plain speed $u(t)$ decreases in the regions where the energy gap $\Delta E$ is small.
If the system at ${t=0}$ is prepared as the ground state of the initial Hamiltonian, this kind of driving can be expected to yield a larger overlap of the evolved state at ${t=T}$ with the final ground state, see \cite{We}.

The metric tensor of Hamiltonian \eqref{HLZ} is diagonal in the spherical parametrization:
\begin{equation}\label{metric2x2}
	g_{\mu\nu}(\boldsymbol{r}) \equiv 
	\left(\begin{array}{ccc}
		g_{rr} & g_{ r\vartheta} & g_{ r\varphi}\\
		g_{ \vartheta r} & g_{\vartheta\vartheta}& g_{ \vartheta\varphi}
		\\
		g_{ \varphi r} & g_{\varphi\vartheta}& g_{ \varphi\varphi}
	\end{array}\right) =
	\left(\begin{array}{ccc}
		0 & 0 & 0\\
		0 & \frac{1}{4}& 0
		\\
		0 & 0& \frac{\sin^{2}\!\vartheta}{4}
	\end{array}\right).
\end{equation}
This metric is singular because of its vanishing radial component, so all curves in the space ${x\times y\times z}$ which have the same radial projection onto the unit sphere cover the same metric length.
This holds in particular for trajectories associated with all the three driving paths discussed here.
Formula~\eqref{metric2x2} directly implies that the semicircular path~$\rm{C}$ also has a constant metric speed ${v= \alpha_0/T}$, the same as for path~$\rm{B}$.
Both these paths can be characterized as \uvo{geometry-inspired}.
In contrast, the ${u={\rm const}}$ linear path~A yields a variable metric speed ${v(t)=4x_0z_0/[T\,\Delta E(\boldsymbol{r}(t))^2]}$ that strongly increases in the small-gap region.

\section{Exact solutions to Schr{\" o}dinger equation}
\label{sec:3}

The time evolution of a coherently driven two-state quantum system is described by the Schr{\" o}dinger equation,
\begin{equation}\label{eq:new2x2Schrodinger}
	\ii\frac{d}{dt}\ket{\psi(t)} = \hat{H}_{\rm LZ}\bigl(\boldsymbol{r}(t)\bigr)\ket{\psi(t)},
	\qquad
	\ket{\psi(t)} = \left(\matrix{a_0(t) \cr a_1(t)\cr }\right),
\end{equation}
where the Planck constant is absorbed into the units of time---see the remark below equation \eqref{HLZ}.
According to the definitions of paths in \eqref{pathA}, \eqref{pathB} and \eqref{pathC}, we set $y =0$ in the Hamiltonian.
For the sake of brevity we hide the dependence of the evolving state vector $\ket{\psi(t)}$ on the path $\boldsymbol{r}(t)$, and particularly on the total driving time $T$ (this dependence will have to be explicitly declared in the forthcoming sections in considerations related to infidelity). Equation \eqref{eq:new2x2Schrodinger} in general represents a coupled system of first-order differential equations. A way to decouple these equations is based on applying the transformation \cite{Rosen,Carr85}
\begin{equation}\label{a_to_Q}
	a_n(t) =\exp\left(\frac{1}{2}\int_{0}^{t}\dot{x}(t')/x(t') dt'\right)Q_n(t), \quad n = 0,1,
\end{equation}
which leads to the following set of differential equations
\begin{eqnarray}	
	\ddot{Q}_0(t) &+ \left(x^2(t) +F^2(t) -\ii \dot{F}(t)\right)Q_0(t) = 0, 
	\label{Sch_gen_imp}
	\\
	Q_1(t) &= -\frac{1}{x(t)} \left( \ii\frac{d}{dt}+ F(t)\right)Q_0(t),
	\label{Q_1_from_1_0}
\end{eqnarray}
where $F(t) \equiv z(t) + \ii\dot{x}(t)/2x(t)$.
The component $Q_0(t)$ is obtained by solving the second-order differential equation \eqref{Sch_gen_imp} and the component $Q_1(t)$ is then calculated from \eqref{Q_1_from_1_0}.
The initial conditions, i.e., the values $Q_0(0)$ and $\dot{Q}_0(0)$, are obtained from the initial state $\ket{\psi(0)}$ expressed from \eqref{gs2x2}, which in our case leads to
\begin{equation}
\left(\matrix{ Q_0(0) \cr Q_1(0)\cr}\right) = \frac{1}{\sqrt{2r_0}}\left(\matrix{\sqrt{r_0-z_0} \cr \sqrt{r_0+z_0} \cr}\right). 
\label{I.C.}
\end{equation}

The method based on \eqref{a_to_Q} assumes that $\dot{x} \neq 0\neq x$.
If $\dot{x}=0\neq x$, which is the case of paths A and B, both differential equations in \eqref{eq:new2x2Schrodinger} can be decoupled directly and equations \eqref{Sch_gen_imp} and \eqref{Q_1_from_1_0} hold with
\begin{equation}
\left(\matrix{ Q_0(t) \cr Q_1(t)\cr}\right)=\left(\matrix{ a_0(t) \cr a_1(t)\cr}\right),
\quad F(t)=z(t)
\qquad{\rm for\ paths\ A,B.}
\label{Q=a}
\end{equation}
This will be used in sections~\ref{subsec_linear} and \ref{sec:caseB} below.
Although for path C we have $\dot{x}\neq 0$, we will see in section~\ref{sec:caseC} that this particular driving can be solved in a simpler way, without reference to formula \eqref{a_to_Q}.

\subsection{Path $\rm{A}$}
\label{subsec_linear}

This driving is the closest to the original Landau-Zener work, where the detuning term varies linearly passing through the avoided-level crossing and the coupling term $x_0$ is constant. In this picture, it turns out convenient to introduce new variables 
\begin{equation}
	X_0 \equiv x_0\sqrt{\frac{T}{2z_0}},\quad Z(t)\equiv z(t)\sqrt{\frac{T}{2z_0}},
\end{equation}
with $z(t)$ defined in \eref{pathA}. These new variables transform~\eref{Sch_gen_imp} into 
\begin{equation}\label{dif_eq_lin_with_change}
	\frac{d^2{\cal Q}}{dZ^2}(Z) + \left(X_0^2 +Z^2 -\ii\right){\cal Q}(Z) = 0,
\end{equation}
where ${\cal Q}(Z(t)) \equiv Q_0(t)$. We relate the above equation to Weber equation, see \ref{app:A}, whose solution is given by the two linearly independent parabolic cylinder functions ${D_{\eta}((1-\ii)Z)}$ and ${D_{\eta}(-(1-\ii)Z)}$, where $\eta = \ii X_0^2/2$. Therefore, with respect to \eqref{Q=a}, the probability amplitudes read as  
\begin{eqnarray}
\label{CaseA_Q0}
\fl \qquad \quad a_0(t) = A_+ D_{\eta}\big((1-\ii)Z(t)\big) + A_- D_{\eta}\big(-(1-\ii)Z(t)\big)\\
\fl \qquad \quad a_1(t) =\frac{1-\ii}{2}X_0\Big[A_+ D_{\eta-1}\big((1-\ii)Z(t)\big)  - A_- D_{\eta-1}\big(-(1-\ii)Z(t)\big)\Big] 
\label{CaseA_Q1},
\end{eqnarray}
where $A_+$ and $A_-$ are some complex coefficients.
After applying Cramer's rule and by using the Wronskian determinant for the parabolic cylinder functions, these coefficients are found to be
\begin{eqnarray}\label{PathA_coef}
\fl	A_{\pm}	&= \frac{\Gamma(1-\eta)}{\sqrt{2\pi}} \Big[
	D_{\eta-1}\big(\mp(1-\ii)Z(0)\big)a_0(0) \pm \frac{1+\ii}{X_0}D_{\eta}\big(\mp(1-\ii)Z(0)\big)a_1(0) \Big],
\end{eqnarray}
where $\Gamma$ is the gamma function and $a_0(0)$, $a_1(0)$ are given by \eref{I.C.}. 

\subsection{Path $\rm{B}$}
\label{sec:caseB}

The time-dependent differential equation \eref{Sch_gen_imp} can be conveniently transformed into an equation depending on angle $\alpha$,
	\begin{equation}\label{dif_eq_geo1_with_change}
		\frac{d^2\tilde{\mathcal{Q}}}{d\alpha^2}(\alpha) +\underbrace{\frac{x_0T}{2\alpha_0}\Big(\frac{x_0T}{2\alpha_0}-\ii\Big)}_{\kappa}\ \sec^2\!\alpha\ \tilde{\mathcal{Q}}(\alpha) = 0,
	\end{equation}
	where $\tilde{\mathcal{Q}}(\alpha(t))\equiv Q_0(t)$. This differential equation can be solved by considering a new variable,
	\begin{equation}
	q(t) = \cos^2\!\alpha(t),
	\end{equation} 
	and by doing the substitution $\tilde{\mathcal{Q}}\mapsto q^{(1-\sqrt{1-4\kappa})/4}w(q)$, which transforms~\eref{dif_eq_geo1_with_change} into Euler's hypergeometric differential equation for the function $w(q)$, with parameters
	 \begin{equation}
	 a = b = \frac{1}{4}\big(1\!-\!\sqrt{1\!-\!4\kappa}\big) = -\ii\frac{x_0T}{4\alpha_0},\quad c = \frac{1}{2}\!+\!2a,
	 \end{equation}
	see~\ref{app:B}.	
	The new variable is bounded by 
	\begin{equation}
		\frac{x_0^2}{r_0^2}\equiv q_0 \leq q \leq 1, \quad t \in[0,T].
	\end{equation}
	Therefore, we can use a special solution to Euler's hypergeometric differential equation near ${q = 1}$ , which we use in~\eref{Q_1_from_1_0} to obtain $Q_1(t)$. With the aid of \eqref{Q=a}, the probability amplitudes are then given by
	\begin{eqnarray}
	\label{case2_Q0}
	\fl
	a_0(t)= q^a(t)\Biggl[ B_1\underbrace{{}_2F_1\Big(a,a;\frac{1}{2}; 1\!-\!q(t)\Big)}_{w_1(q(t))}
	+\,B_2\underbrace{\sin\alpha(t)\, {}_2F_1\Big(\frac{1}{2}\!+\!a,\frac{1}{2}\!+\!a;\frac{3}{2}; 1\!-\!q(t) \Big)}_{w_2(q(t))}\Biggr],
\\
\fl
	a_1(t) = -\ii\frac{q^{a}(t)}{x_0}\Big[B_1\dot{w}_1\big(q(t)\big)+B_2\dot{w}_2\big(q(t)\big)\Big]
	\label{case2_Q1}
	\end{eqnarray}
	where $q^a(t)$ is a time-dependent phase, ${}_2F_1$ is the hypergeometric function, $B_1, B_2$ are some complex coefficients, and the total time derivative of functions $w_1,w_2$ is explicitly given by $\dot{w}_n = \frac{d w_n}{d q}\frac{d q}{d \alpha}\,\dot{\alpha}$. 
	The coefficients can be again determined from the initial conditions, which lead to
	\begin{eqnarray}\label{PathB_coef}
		B_1 = - & 2aq_0^a\Big( a_0(0)\frac{x_0z_0}{ar_0^2}w'_2(q_0) + a_1(0)w_2(q_0)\Big), \\
		B_2	=   & 2aq_0^a\Big( a_0(0)\frac{x_0z_0}{ar_0^2}w'_1(q_0) + a_1(0)w_1(q_0)\Big),
	\end{eqnarray}
	where we use notation $w'_n\equiv \frac{d w_n}{d q}$.

\subsection{Path $\rm{C}$}
\label{sec:caseC}

The motion along the arc segment involved in path C keeps a constant angular velocity.
This implies that we can describe the system in a rotating frame, which connects to the static system through the time-dependent unitary transformation
\begin{equation}\label{transformation}
	\ket{\psi(t)} = \ee^{\ii\hat{\sigma}_y\alpha(t)/2}\ket{\tilde{\psi}(t)}.
\end{equation}
This transforms the original Hamiltonian (\ref{HLZ}) into the time-independent Hamiltonian
\begin{equation}\label{time_independent_Hamiltonian}
	\hat{\tilde{H}} = \ee^{-\ii\hat{\sigma}_y\alpha(t)/2}\hat{H}_{\rm LZ}\bigl(\boldsymbol{r}(t)\bigr)\ee^{\ii\hat{\sigma}_y\alpha(t)/2} = -r_0 \hat{\sigma}_x  + \frac{\alpha_0}{T}\hat{\sigma}_y.
\end{equation}
From the solution to the time-independent Schr{\" o}dinger equation, it immediately follows that the evolving state vector is given by
\begin{equation}\label{Scho_sol_case3}
	\ket{\psi(t)} =\ \ \ee^{\ii\hat{\sigma}_y\alpha(t)/2}\ee^{-\ii\hat{\tilde{H}}t}\ee^{\ii\hat{\sigma}_y\alpha_0/2}\ket{\psi(0)},
\end{equation}
or in the matrix form,
\begin{equation}
\fl
	\left(\!\matrix{ a_0(t) \cr a_1(t)\cr}\!\right) =
	\left(\!\!\begin{array}{cc}
		A(t) \cos \frac{\alpha_0}{2} \!-\! B(t) \sin \frac{\alpha_0}{2}  & A(t) \sin \frac{\alpha_0}{2} \!+\! B(t) \cos \frac{\alpha_0}{2}
		\\
		-A(t)^* \sin \frac{\alpha_0}{2} \!-\! B(t)^* \cos \frac{\alpha_0}{2}& A(t)^* \cos \frac{\alpha_0}{2} \!-\! B(t)^* \sin \frac{\alpha_0}{2}
	\end{array}\!\!\right)	
	\left(\!\matrix{ a_0(0) \cr a_1(0)\cr}\!\right),
	\label{Cmatrix}
\end{equation}
where the star represents the complex conjugation and
\begin{eqnarray}
	A(t) &=& \cos\frac{\pi t}{\tau}\cos\frac{\alpha(t)}{2}+\frac{\tau}{\pi}\Bigl(\frac{\alpha_0}{T}\!+\!\ii r_0 \Bigr)\sin\frac{\pi t}{\tau}\sin\frac{\alpha(t)}{2}, \\
	B(t) &=& \cos\frac{\pi t}{\tau} \sin\frac{\alpha(t)}{2}-\frac{\tau}{\pi}\Bigl(\frac{\alpha_0}{T}\!-\!\ii r_0 \Bigr)\sin\frac{\pi t}{\tau}\cos\frac{\alpha(t)}{2},
\end{eqnarray}
with $\tau = \pi/\sqrt{r_0^2 + \alpha_0^2/T^2}$.
The initial state in equation \eqref{Cmatrix} is taken from \eqref{I.C.}.

\section{Instantaneous infidelity} 
\label{sec:4}

\begin{figure}[tp]
\begin{adjustbox}{right}
\includegraphics[width=.88\textwidth]{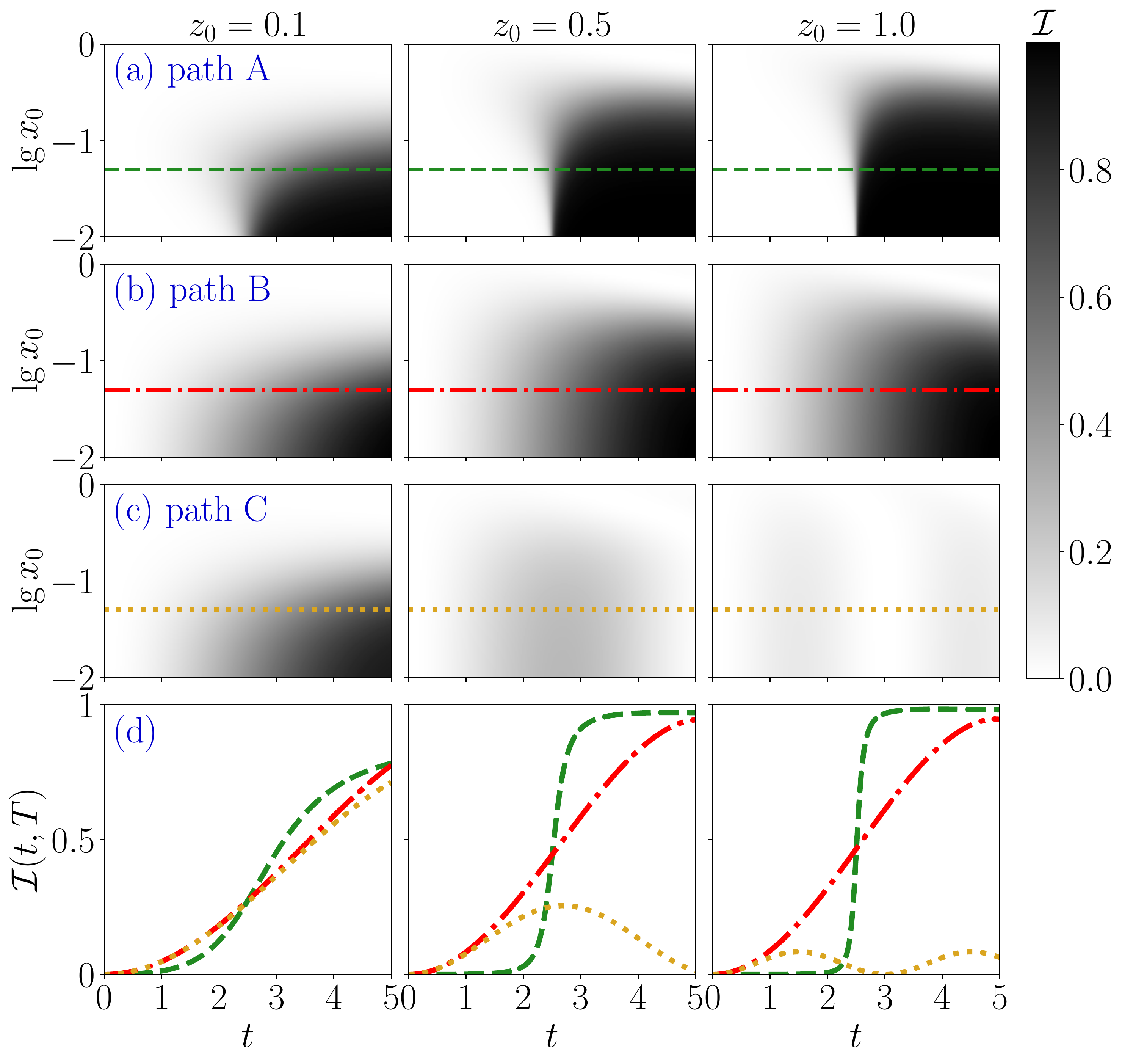}
\end{adjustbox}
\caption{Instantaneous infidelity for the three drivings with various choices of parameters $x_0$ and $z_0$ and fixed total time $T=5$.
The upper three rows, (a) for path A, (b) for path B, and (c) for path C, show density plots of the infidelity as a function of $t$ and $x_0$ for $z_0=0.1$ (left), $z_0=0.5$ (middle) and $z_0=1.0$ (right). Row (d) depicts the infidelity dependencies along the $x_0=0.05$ cut of the upper plots.
The cut is indicated in rows (a)--(c) by horizontal lines whose type distinguishes the driving type in panel (d).}
	\label{fig:Figure3}
\end{figure}

In the previous section we obtained exact expressions for $a_0(t)$ and $a_1(t)$, the probability amplitudes, for all three paths A, B and C. We proceed to use these solutions for the purpose of target state preparation. The system starts in the ground state of the initial Hamiltonian $H_{\rm LZ}(\boldsymbol{r}(0))$ and shall be delivered to the ground state of the final Hamiltonian $H_{\rm LZ}(\boldsymbol{r}(T))$, which is our target state \cite{Zana06}.
With respect to these settings we introduce the instantaneous infidelity 
\begin{equation}\label{def_Infidelity_official}
	\mathcal{I}(t,T) \equiv 1- \underbrace{\big|\left\langle E_{0}(\boldsymbol{r}(t)) \middle\vert \psi(t)\right \rangle\big|^2}_{\mathcal{F}(t,T)}= \big|\left\langle E_{1}(\boldsymbol{r}(t)) \middle\vert \psi(t)\right \rangle\big|^2.
\end{equation}
which represents the time-dependent probability that the system is \emph{not} found in the immediate ground state at the parameter point $\boldsymbol{r}(t)$. Here we explicitly declare the dependence of infidelity on the total driving time $T$, the upper limit of variable $t$, which is hided in the notation of $\ket{\psi(t)}\equiv(a_0(t),a_1(t))^{\rm T}$. The infidelity is obviously related to instantaneous fidelity $\mathcal{F}(t,T)=1-\mathcal{I}(t,T)$. Inserting the immediate state $\ket{E_1(\boldsymbol{r})}$ from \eqref{es2x2} into \eqref{def_Infidelity_official}, we obtain
\begin{equation}\label{def_Infidelity_AB}
	\mathcal{I}(t,T) = \frac{r(t)-z(t)\bigl(1-2|a_1(t)|^2\bigr) -2x(t){\rm{Re}}\bigl[a_0(t)a_1^{*}(t)\bigr] }{2r(t)}.
\end{equation}
Whilst the instantaneous infidelity $\mathcal{I}(t,T)$ does not have direct consequences for the final infidelity $\mathcal{I}(T,T)$, which measures the deviation of the final state $\ket{\psi(T)}$ from the target state $\ket{E_{0}(\boldsymbol{r}(T))}$, its $t\in(0,T)$ dependence provides essential insight into the mechanisms generating departures from the desired transitionless evolution during the driving \cite{Farh01}. It was also used in this sense in \cite{We}.

The instantaneous infidelity for all three driving paths A, B and C evaluated for various choices of parameters $x_0$ and $z_0$ with a fixed total driving time $T=5$ are shown in figure~\ref{fig:Figure3}. For drivings A and B, which follow the same line in the parameter space, at time $t = T/2$ passing in the minimal distance $x_0$ to the exact $r=0$ level crossing, we observe (in the two upper rows) an increase of infidelity $\mathcal{I}_{\rm{A}}(t,T)$ and $\mathcal{I}_{\rm{B}}(t,T)$ in the middle of the time interval. This is not surprising since the nonadiabatic effects play the most important role in the small-gap region. However, these effects are apparently reduced for the geometry-inspired path B, which minimizes the speed $u$ near the crossing and therefore exhibits a softer increase of infidelity. The increase of parameter $z_0$ (in columns from left to right) for a fixed $T$ increases the overall speed average, hence also the speed near the avoided crossing, and therefore leads to a sharpening of the infidelity increase in the mid-time region. 
Let us note that the substitution of solutions from sections~\ref{subsec_linear} and \ref{sec:caseB} into \eqref{def_Infidelity_AB} yields analytic expressions of $\mathcal{I}_{\rm{A}}(t,T)$ and $\mathcal{I}_{\rm{B}}(t,T)$, but these expressions are too long and we do not show them here.

For path C, the instantaneous infidelity $\mathcal{I}_{\rm{C}}(t,T)$ in the lower row of figure~\ref{fig:Figure3} shows a different behavior than the infidelity of paths A and B. In this case, using the results of section~\ref{sec:caseC}, we can derive a simple formula
\begin{equation}\label{Infidelity_geodesic}
	\mathcal{I}_{\rm{C}}(t,T) = \underbrace{\frac{\alpha_0^2}{r_0^2T^2 + \alpha_0^2}}_{\bar{\mathcal{I}}_{\rm{C}}} \sin^2
	\Biggl(t\underbrace{\sqrt{r_0^2 +\frac{\alpha_0^2}{T^2}}}_{\pi/\tau}\Biggr).
\end{equation}
It predicts oscillations of infidelity between minimal and maximal values $\mathcal{I}_{\rm{C}}=0$ and $\mathcal{I}_{\rm{C}}=\bar{\mathcal{I}}_{\rm{C}}<1$, respectively,  with period $\tau$.
The exact zeros of \eqref{Infidelity_geodesic} appear at times
\begin{equation}\label{t_n_zeros}
t=t_{k}=k\tau
\qquad\text{for } k=0,1,2,\dots,\left\lfloor\frac{T}{\tau}\right\rfloor,
\end{equation}
where $\lfloor x\rfloor$ denotes the floor function.
This is indeed observed in figure~\ref{fig:Figure3}, where we see the first and second oscillation of $\mathcal{I}_{\rm{C}}(t,T)$ as $z_0$ increases from 0.1 to 1.
Exact zeros of instantaneous infidelity appear also for paths A and B, though much less regularly than for the path C.
The equality ${\mathcal{I}(t,T)=0}$ for an arbitrary driving of the two-level system is equivalent to the condition
\begin{equation}
\frac{a_1(t)}{a_0(t)}=\frac{x(t)+\ii\, y(t)}{z(t)+r(t)}.
\end{equation} 
For paths A and B this condition is satisfied in isolated time instants near the final time $T$ (it turns out that this happens only if $T$ is long enough). Let us stress that zeros of infidelity do not generically exist in systems with Hilbert-space dimensions $d>2$ since in such systems the ${\mathcal{I}(t,T)=0}$ condition requires to simultaneously satisfy multiple constraints (the evolved state must have zero overlaps with all excited states). 

\begin{figure}[tp]
\begin{adjustbox}{right}
\includegraphics[width=\textwidth]{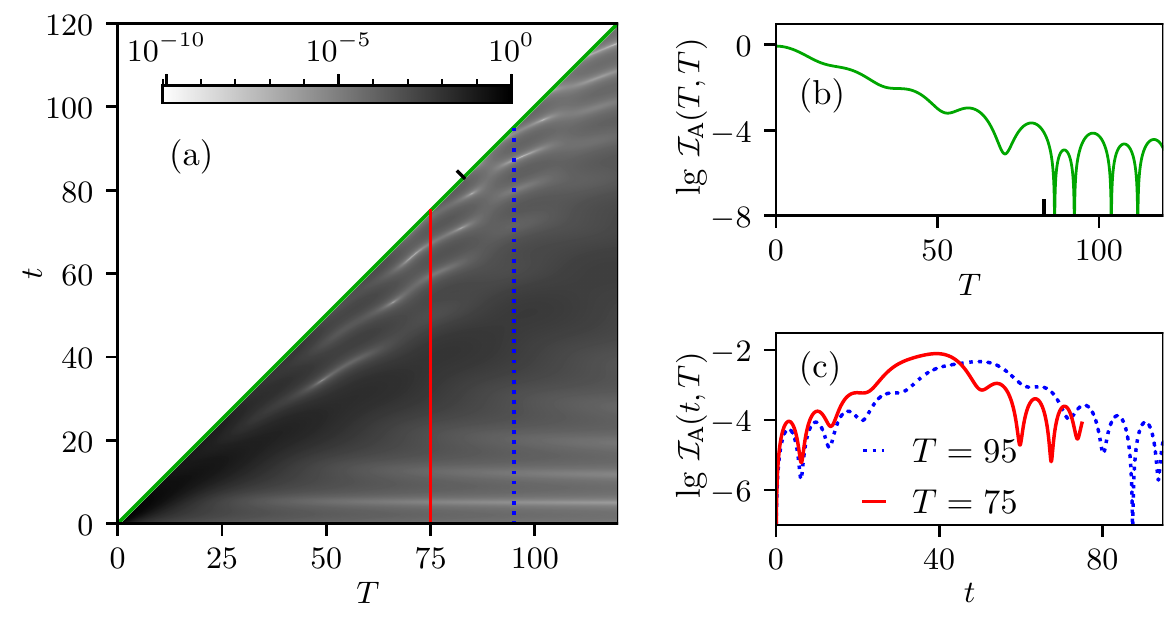}
\end{adjustbox}
\caption{Instantaneous infidelity $\mathcal{I}_{\rm{A}}(t,T)$ for path A as a function of time $t\in[0,T]$ and the total driving time $T$ for $x_0=0.2$, $z_0=0.5$. The density plot in panel (a) depicts the whole dependence in the $t\times T$ plane, panel (b) shows the final infidelity $\mathcal{I}_{\rm{A}}(T,T)$, and panel (c) $\mathcal{I}_{\rm{A}}(t,T)$ for $T=75$ and 95; these cuts are indicated in panel (a) by vertical lines. The tick on the diagonal line in panel (a) and on the horizontal axis in panel (b) represents the crossover time $T_{\rm c}$ between the LZ and APT regimes discussed in section~\ref{sec:5}. We see that above $T_{\rm c}$ the gorges of $\mathcal{I}_{\rm{A}}(t,T)$ in panel (a) lean towards the $t=T$ line, so oscillations of infidelity reaching nearly zero values appear close to the end of driving; cf. the $T=95$ curve in panel (c).}
	\label{fig:birdA}
\end{figure}

\begin{figure}[tp]
\begin{adjustbox}{right}
\includegraphics[width=\textwidth]{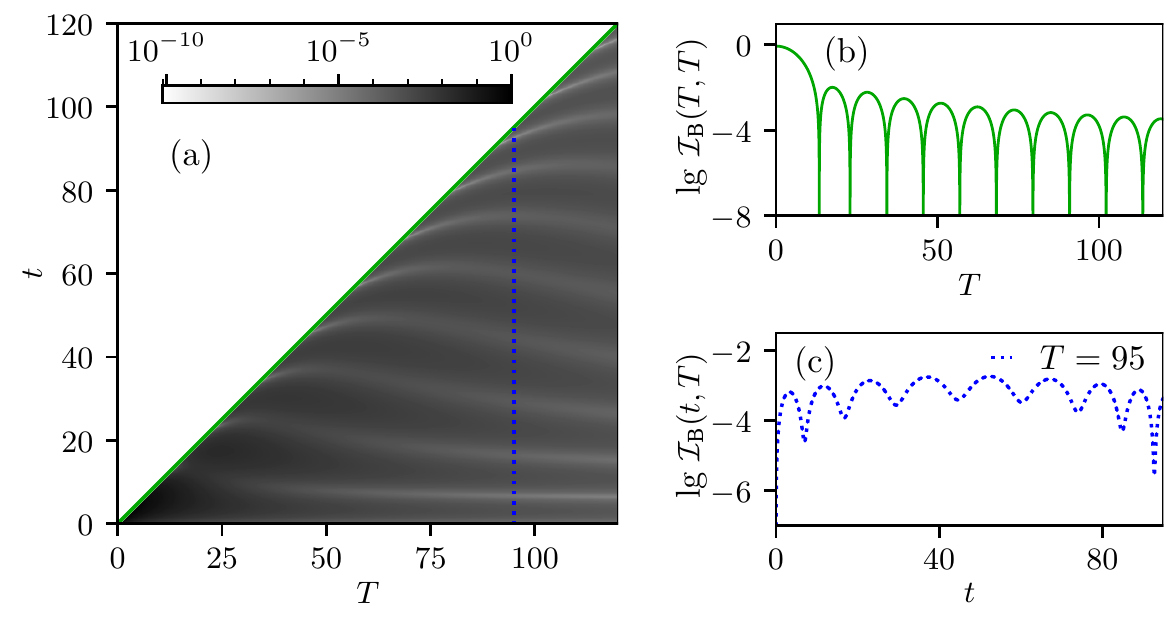}
\end{adjustbox}
\caption{Instantaneous infidelity $\mathcal{I}_{\rm{B}}(t,T)$ for driving B as a function of $t$ and~$T$. Parameter values and the organization of the figure are the same as in figure~\ref{fig:birdA}. Since there is no crossover time for drivings B and C, we plot only the cut of $\mathcal{I}_{\rm{B}}(t,T)$ for $T=95$.}
	\label{fig:birdB}
\end{figure}

\begin{figure}[tp]
\begin{adjustbox}{right}
\includegraphics[width=\textwidth]{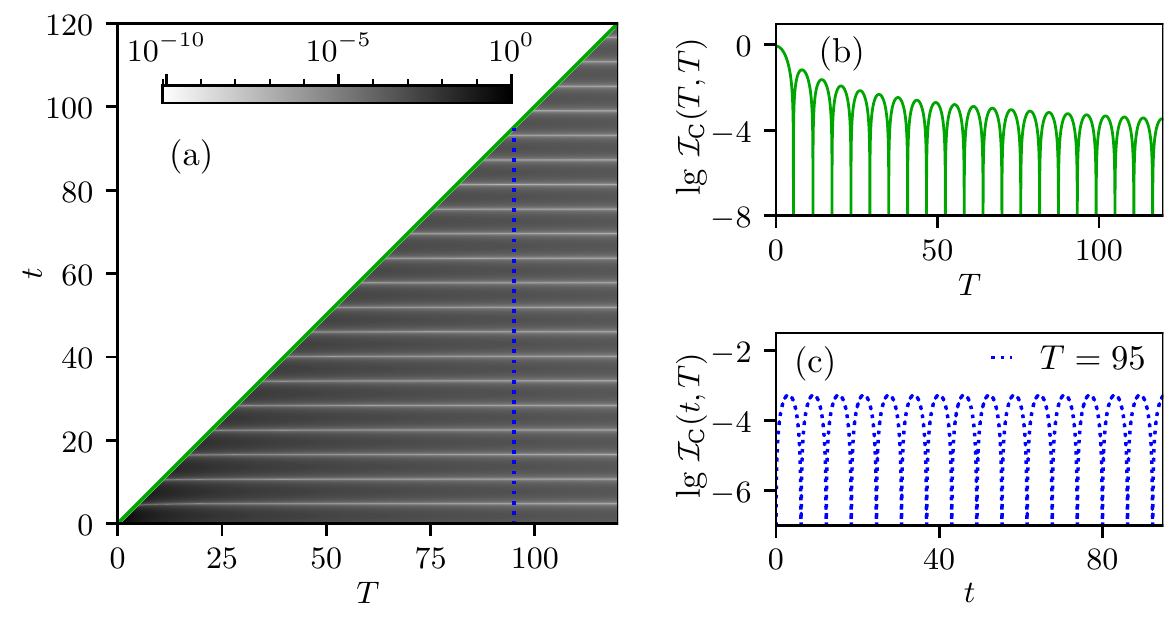}
\end{adjustbox}
\caption{Instantaneous infidelity $\mathcal{I}_{\rm{C}}(t,T)$ for driving C for the same parameters as in figures~\ref{fig:birdA} and \ref{fig:birdB}.
	}
	\label{fig:birdC}
\end{figure}

The instantaneous infidelity for all three paths for fixed parameters ${x_0=0.2}$ and ${z_0=0.5}$ and variable driving time $T$ are depicted in figures~\ref{fig:birdA} (path A), \ref{fig:birdB} (path B), and~\ref{fig:birdC} (path C).
These may be seen as bird's flight views of our analytic solutions (the resulting infidelities) in the plane $t\times T$.

\section{Final infidelity} 
\label{sec:5}

The well known closed expression for the final infidelity in the two-level system is the LZ formula (sometimes called Landau-Zener-St{\" u}ckelberg-Majorana formula) \cite{Land32,Zene32,Stue32,Majo32}. As explained in section~\ref{sec:2}, this formula is derived for a system initiated at time $t\to-\infty$ in the ground state $\ket{E_0(\boldsymbol{r}_{\rm I})}$ of the LZ Hamiltonian at asymptotic parameter point $\boldsymbol{r}_{\rm I}=\lim_{t\to-\infty}\boldsymbol{r}(t)$ and evolving with a constant velocity $\boldsymbol{u}$ along the line ${\boldsymbol{r}(t)=\boldsymbol{r}(0)+\boldsymbol{u}t}$ to the $t\to +\infty$ asymptotic point $\boldsymbol{r}_{\rm F}=\lim_{t\to+\infty}\boldsymbol{r}(t)$.
The final infidelity of the target state $\ket{E_0(\boldsymbol{r}_{\rm F})}$ reads as
\begin{equation}
\label{LZformula}
	\mathcal{I}^{\rm LZ} = \ee^{-\pi |\boldsymbol{r}(0)|^2/u}.
\end{equation}
If $|\boldsymbol{r}(0)|^2/u$ is large enough, the driving is adiabatic, yielding $\mathcal{I}_{\rm LZ}\approx 0$. For our drivings in the restricted parameter range the LZ formula \eqref{LZformula} is generally not valid, but it may provide a good approximation of the final infidelity for path A within a certain driving-time domain. 

\begin{figure}[t]
\begin{adjustbox}{right}
		\includegraphics[width=\textwidth]{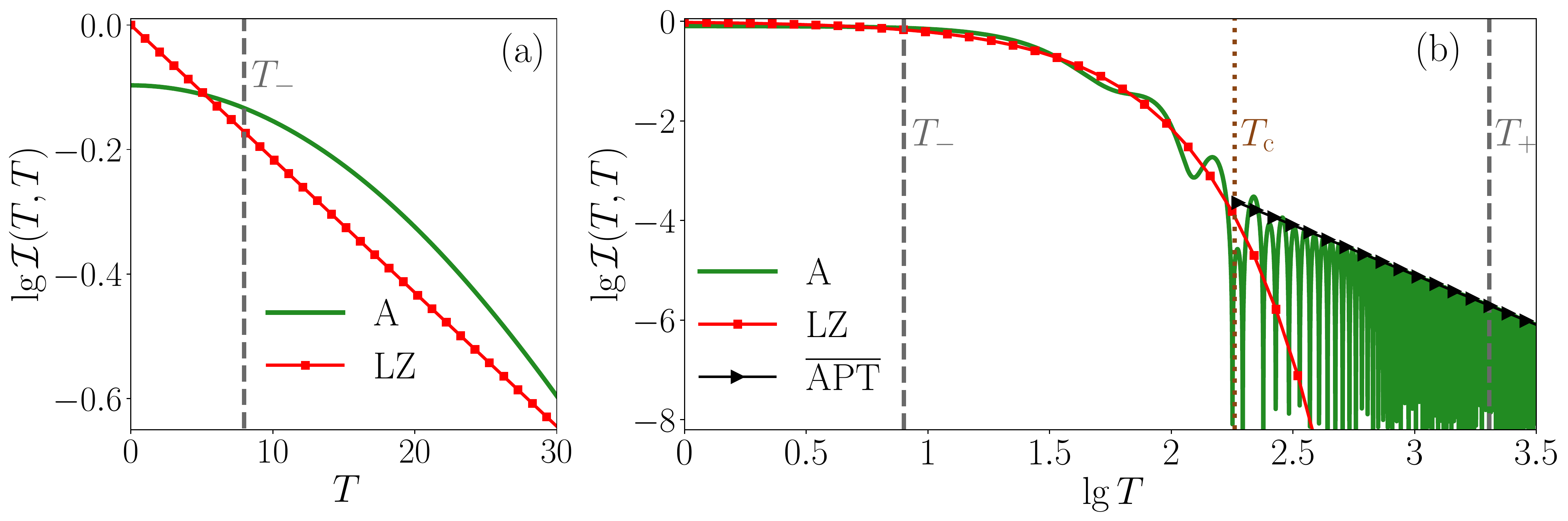}
\end{adjustbox}
\caption{The dependence of the final infidelity for path A on time $T$.
Panel~(a): The lin-log plot visualizing the short- and medium-$T$ domain, with the exact infidelity (the solid curve) and the LZ approximation \eqref{LZI} (the tilted line). Panel~(b): The log-log plot visualizing the medium- and long-$T$ domain, with the exact infidelity (the solid oscillating curve), LZ approximation  (the descending curve), and the upper envelope of the APT approximation \eqref{APT:case_A} (the tilted line). The times $T_{\pm}$ and $T_{\rm c}$ from \eqref{Tpm} and \eqref{Lambert} are marked. The parameters are ${x_0 = 0.063}$, ${z_0 = 0.126}$.}
\label{fig:Figure7}
\end{figure}

As shown in \ref{app:A}, the LZ formula represents the lowest-order term of the asymptotic expansion of the exact solution to driving A in terms of the parabolic cylinder functions. The expansion is applicable for $z_0/x_0>1/\sqrt{2}$ if the total driving time $T$ satisfies the condition 
\begin{equation}\label{Tpm}
T_-\lesssim T\lesssim T_+,\qquad T_{\pm}=\frac{8z_0^3}{x_0^4}\left(1\pm\sqrt{1-\frac{x_0^4}{4z_0^4}}\right).
\end{equation}
Only somewhere within these bounds we can write
\begin{equation}
\mathcal{I}_{\rm A}(T,T)\approx\mathcal{I}_{\rm A}^{\rm LZ}(T) =\ee^{-\pi x_0^2T/2z_0},
\label{LZI}
\end{equation}
bearing in mind that in the actual interval of validity of the LZ approximation the exact solution $\mathcal{I}_{\rm A}(T,T)$ oscillates around $\mathcal{I}_{\rm A}^{\rm LZ}(T)$.
On the long-time side, the LZ approximation fails for driving times $T$ reaching a special value ${T_{\rm c}<T_+}$, which represents a crossover to a non-LZ regime of driving discussed below, see \eqref{Lambert}.
On the short-time side, the existence of the limit $T_-$, below which the exact infidelity also deviates from the LZ approximation, is a direct consequence of bounds in the parameter space. In  the LZ approach, the parameter distance of the initial and target ground states is infinite, so these states are orthonormal. In the bounded space, in contrast, we have ${\scal{E_0(\boldsymbol{r}_{\rm F})}{E_0(\boldsymbol{r}_{\rm I})}\neq 0}$, hence the diabatic limit $T\to 0$ for all drivings A, B and C yields
\begin{equation}
\lim_{T\to 0}\mathcal{I}(T,T)=|\scal{E_1(\boldsymbol{r}_{\rm F})}{E_0(\boldsymbol{r}_{\rm I})}|^2=\frac{z_0^2}{r_0^2}\leq 1.
\end{equation}
These conclusions are demonstrated, for a particular choice of parameters, in figure~\ref{fig:Figure7}. Note that for $z_0/x_0\leq 1/\sqrt{2}$ the interval defined in \eqref{Tpm} disappears and the LZ approximation fails everywhere.

As discussed in reference~\cite{We}, the final infidelity for a general driving path with sufficiently long driving time $T$ is reliably determined from the adiabatic perturbation theory (APT) \cite{Ortiz} (for a brief summary of the APT see \ref{app:C}). In the APT regime, the dependence of the final infidelity on $T$ shows different features than in the LZ regime, in particular, ${\cal I}(T,T)$ is well approximated by an algebraic expansion in terms of powers $T^{-p}$ with ${p\geq 2}$. In a generic case, which includes all the three driving paths considered here, the leading-order contribution has $p=2$, so we can write 
\begin{equation}
 {\cal I}(T,T)\approx{\cal I}^{\rm APT}(T) =\frac{{\cal I}^{(2)}(T)}{T^{2}},
 \label{APTI}
\end{equation}
where ${\cal I}^{(2)}(T)$ is a coefficient given by a general expression from \cite{We}. For paths A, B and C we obtain:
\begin{eqnarray}
	\mathcal{I}_{\rm{A}}^{\text{APT}}(T) &=\underbrace{\frac{x_0^2z_0^2}{r_0^6T^2}}_{\overline{{\cal I}}{}_{\rm A}^{\rm APT}(T)}\sin^2 \left(\frac{T}{2}\left[r_0 + \frac{x_0^2}{2z_0}\ln\frac{r_0+z_0}{r_0-z_0}\right]\right), 
	\label{APT:case_A} \\
	\mathcal{I}_{\rm{B}}^{\text{APT}}(T) &=\underbrace{\frac{\alpha_0^2}{r_0^2T^2}}_{\overline{{\cal I}}{}_{\rm B}^{\rm APT}(T)}\sin^2 \left(\frac{T}{2} \frac{x_0}{\alpha_0}\ln\frac{r_0+z_0}{r_0-z_0}\right), 
	\label{APT:case_B}\\
	\mathcal{I}_{\rm{C}}^{\text{APT}}(T) &=\underbrace{\frac{\alpha_0^2}{r_0^2T^2}}_{\overline{{\cal I}}{}_{\rm C}^{\rm APT}(T)}\sin^2 \big(Tr_0\big).
	\label{APT:case_C}
\end{eqnarray}

As seen in \eqref{APT:case_A}--\eqref{APT:case_C}, the coefficient ${\cal I}^{(2)}(T)$ in \eqref{APTI} carries an oscillatory dependence on $T$. Let us note that a residual dependence of coefficients ${\cal I}^{(p)}(T)$ in the APT expansion on the expansion variable $T^{-1}$ is a general feature of the method \cite{Ortiz}, which does not deteriorate its performance because of the bounded, fast oscillating character of this dependence.  One can evaluate the maximal value of ${\cal I}^{(2)}(T)$, determining an upper envelope $\overline{{\cal I}}{}^{\rm APT}(T)$ of the expression \eqref{APTI}. An example for path A is seen in panel (b) of figure~\ref{fig:Figure7}, which depicts a log-log dependence of the exact final infidelity $\mathcal{I}_{\rm{A}}(T,T)$ on the driving time. We observe that for smaller values of $T$ the exact dependence conforms with \eqref{LZI}, while for larger values of $T$ it is consistent with formula \eqref{APT:case_A}. The crossing of the LZ approximation $\mathcal{I}_{\rm A}^{\rm LZ}(T)$ with the APT average $\overline{{\cal I}}{}^{\rm APT}_{\rm A}(T)/2$ roughly demarcates a crossover between the LZ and APT regimes. The crossing happens at
\begin{equation}
T=T_{\rm c}\equiv-\frac{4z_{0}}{\pi x_{0}^2}\ W_{-1}\biggl(-\frac{\pi x_{0}^3}{4\sqrt{2}r_{0}^3}\biggr),
\label{Lambert}
\end{equation}
where $W_{-1}$ is the Lambert function defining the ${w\leq-1}$ branch of solutions to the equation ${w\ee^w=a}$ in the interval ${-\ee^{-1}\leq a<0}$. The solution exists only for $z_0/x_0\geq 0.562$, which is close to the condition for the existence of the LZ approximation resulting from \eqref{Tpm}, otherwise $T_{\rm c}$ is not defined and the onset of the APT regime has a moderate character \cite{We}. Of course, this is true also for driving paths B and C, which show no LZ regime and therefore no crossover time $T_{\rm c}$ for any parameter choice.

\begin{figure}[t!]	
		\begin{adjustbox}{right}
		\includegraphics[width=.8\textwidth]{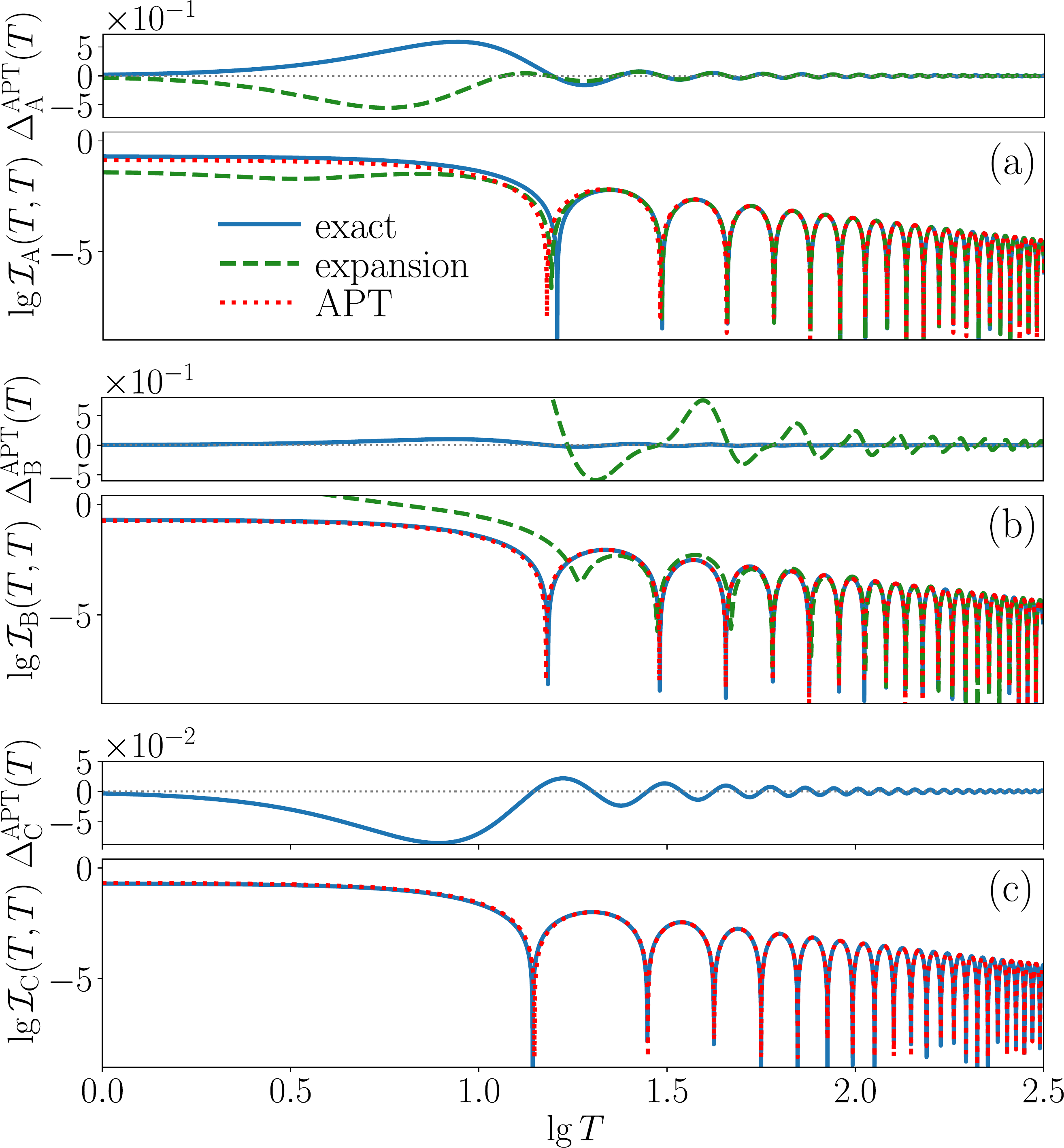}
		\end{adjustbox}
	\caption{A comparison of exact final infidelities (full curves) for all paths  with their APT approximations (dotted curves) and with asymptotic expansions of exact solutions A and B (dashed curves). The paths A, B and C are presented in panels (a), (b) and (c), respectively. The lower plot in each panel depicts the log-log dependence of infidelity of the final time, while the upper plot represents a relative deviation of the exact or expanded solution from the APT solution: $\Delta^{\rm APT}_{\bullet}(T)\!\!=[{\cal I}_{\bullet}(T,T)-{\cal I}_{\bullet}^{\rm APT}(T)]/\overline{{\cal I}}{}_{\bullet}^{\rm APT}(T)$ with $\bullet$\,=A,B,C. The asymptotic expansions included to the comparison in panels (a) and (b) are described in \ref{app:A} and \ref{app:B}, respectively. The parameters are ${x_0=0.2}$ and  ${z_0 = 0.1}$.}
	\label{fig:Figure8}
\end{figure} 

For paths A and B we observe (cf. figures~\ref{fig:birdA}, \ref{fig:Figure7} and \ref{fig:Figure8}) that the transition to the APT regime is usually (for most of parameter choices) signaled by the onset of large-amplitude oscillations of instantaneous and final infidelities with time.  This is also seen in the expressions \eqref{APT:case_A} and \eqref{APT:case_B}, in which the coefficient ${\cal I}^{(2)}(T)$ is proportional to a single squared sine function [the same holds also for path C, but in that case large infidelity oscillations characterize the full solution in \eqref{Infidelity_geodesic}]. Let us point out that oscillatory time dependencies are inherent in general solutions to the basic differential equation~\eqref{Sch_gen_imp}, which can give rise to rather irregular patterns of infidelity oscillations (cf.\,\cite{We}). Moreover, the present observation seems to be restricted only to the two-level model, whereas models in higher dimensional Hilbert spaces will generically yield more complex forms of coefficients ${\cal I}^{(2)}(T)$ and therefore more ambiguous relations between the onset of oscillations and start of the APT regime.

We stress that the APT regime of the dependence ${\cal I}(T,T)$ must be present in all finite-time solutions of the driving problem if $T$ is long enough (although for some driving protocols the leading-order contribution can be shifted to some $p>2$ terms of the APT expansion \cite{We}). Comparison of the present exact analytic solutions with the ${p=2}$ leading-order APT prediction is shown in figure~\ref{fig:Figure8}. For path C the relation is straightforward since we can immediately see that the leading-order term obtained by expanding the exact solution \eqref{Infidelity_geodesic} for ${t=T}$ in powers of $T^{-1}$ coincides with the APT expression \eqref{APT:case_C}. For paths A and B, the comparison is much more difficult to obtain in an analytic form, but figure~\ref{fig:Figure8} shows it pictorially. The parabolic cylinder functions and hypergeometric functions involved in explicit solutions for paths A and B (sections~\ref{subsec_linear} and \ref{sec:caseB}) can be approximated by asymptotic expansions, which are discussed in \cite{Olver} and \cite{Watson}, and outlined in \ref{app:A} and \ref{app:B}, respectively. These approximations (which are very hard to work with) are included into the graphical comparisons in figure~\ref{fig:Figure8}, but we see that they certainly do not do a better job than the simple APT approximation. So we can conclude here that the APT represents the easiest way to approximate the exact solutions for large enough driving time $T$.

\section{Summary}
\label{sec:6}

We have investigated quantum driven dynamics in a restricted parameter space of a two-level system. The presence of parameter bounds is in contrast to the Landau-Zener analytic solution to the driving through a region of avoided crossing of two energy levels, which was derived under the assumption of asymptotic initial and final states. Considering the initial and final parameter points $\boldsymbol{r}_{\rm I}$ and $\boldsymbol{r}_{\rm F}$ placed in a finite distance ${|\boldsymbol{r}_{\rm I}|=|\boldsymbol{r}_{\rm F}|}$ from the ${\boldsymbol{r}=0}$ diabolic point and initializing the system in the ground state at $\boldsymbol{r}_{\rm I}$, we have presented three analytic solutions to the time-dependent Schr{\" o}dinger equation for different ${\boldsymbol{r}_{\rm I}\to\boldsymbol{r}_{\rm F}}$ driving paths: (a) path A was a line between the initial and final points (through the region of an avoided crossing of both levels) passed with constant plain speed $u$ in the parameter space (variable metric speed $v$), (b) path B was the same line passed with constant metric speed $v$ (variable plain speed $u$, slowing down in the avoided-crossing region), (c) path C was an arc segment that bypasses the avoided-crossing region with constant plain and metric speeds $u$ and $v$.

With the aid of the three analytic solutions, we have analyzed the dependence of instantaneous and final infidelity (probability of finding the system in the excited state) on the time and initial/final points. We have demonstrated significant complexity of solutions to the restricted-space driving problem which is manifested already in the most trivial two-level case. The solutions describe regular (path C) or irregular (paths A and B) oscillations of infidelity, which leads to a possibility of completely transitionless drivings for some particular parameter values and driving times (this is general feature of driving in two-level systems, which however does not extend to higher dimensions). Only path A may show a time interval where the final infidelity is well approximated by the Landau-Zener formula. On the other hand, all paths exhibit a regime of long driving times $T$, in which the infidelity is governed by the leading-order ($\propto T^{-2}$) contribution to the expansion obtained within the adiabatic perturbation theory. The APT turns out to be a simple and versatile method (much more effective than the asymptotic expansions of our exact solutions) which reliably approximates exact solutions (including zeros of infidelity) in the case of slow driving.

Let us finally note that the solutions presented here are certainly not the only bounded-space analytic solutions to the two-level driving problem that can be found. Numerous other solutions associated with different driving paths must necessarily exist. The present work is intended as a case study demonstrating general features of such solutions. It can be beneficial in testing new approximation methods in the theory of driven dynamics, as well as in predicting outcomes of concrete experiments.

\section*{Acknowledgments}
We acknowledge financial support of the Czech Science Foundation (Grant No. 20-09998S) and the Charles University in Prague (UNCE/SCI/013).

\appendix

\section{Parabolic cylinder functions}
\label{app:A}

The parabolic cylinder function $D_{\eta}(\xi)$ \cite{HTF} is a solution to the Weber differential equation, 
\begin{equation}\label{appendix1:weber_dif_eq}
	\frac{d^2 }{d\xi^2}D_{\eta}(\xi) + \Big(\eta + \frac{1}{2}- \frac{\xi^2}{4}\Big)D_{\eta}(\xi) = 0
\end{equation}
Using Whittaker's notation, all solutions are $D_{\eta}(\pm\xi) $ and $D_{-\eta-1}(\pm\ii\xi)$. 
If $\eta$ is not an integer then $D_{\eta}(\xi)$ and $D_{\eta}(-\xi)$ are linearly independent, whose Wronskian relation is given by, 
\begin{equation}\label{appendix1:Wronskian}
	W_{\xi}\big\{D_{\eta}(\xi),D_{\eta}(-\xi)\big\} = \frac{\sqrt{2\pi}}{\Gamma(-\eta)},
\end{equation}
where the derivative is explicitly given by
\begin{equation} \label{appendix1:derivative}
	\frac{d}{d\xi} D_{\eta}(\xi) = \eta D_{\eta-1}(\xi) - \frac{1}{2}\xi D_{\eta}(\xi).
\end{equation}

The asymptotic expansion for fixed $\eta$ and large $|\xi|$, provided that $|\arg \xi| < 3\pi/4$, is given by
\begin{equation}\label{appendix1:expansion}
	D_{\eta}(\xi) = \xi^{\eta} \ee^{-\xi^2/4}\left[\sum_{k=0}^{N} \frac{\left(-\frac{\eta}{2}\right)_k \left(\frac{1-\eta}{2}\right)_k}{k! \left(-\frac{\xi^2}{2}\right)_k} + \mathcal{O}\big(|\xi|^{-2(N+1)}\big)\right],
\end{equation}
where ${(x)_{k} = \Gamma(x+k)/\Gamma(x)}$  is the Pochhammer symbol.
For other values of $\arg \xi$ the above expansion must be used with a connection formula. In particular we consider
\begin{equation}\label{appendix1:connection}
	D_{\eta}(\xi) = \ee^{\eta\pi\ii}D_{\eta}(-\xi) + \frac{\sqrt{2\pi}}{\Gamma(-\eta)}\ee^{(\eta+1)\pi\ii/2} D_{-\eta-1}(-\ii\xi).
\end{equation}
In section \ref{sec:5} the expansion \eqref{appendix1:expansion} was used in order to get a closed expression for the final infidelity for ${\eta = \ii x_0^2T/4z_0}$ and ${\ii x_0^2T/(4z_0)-1}$ with ${\xi = \pm (1-\ii)\sqrt{z_0T/2}}$.
Provided that ${|\xi| > |\eta|}$, that is, ${z_0T > 1+ x_0^4T^2/(16z_0^2)}$, which is equivalent to \eqref{Tpm}, the LZ formula is obtained as the final infidelity at the lowest order, ${k=0}$. When the conditions for the validity of expansion \eqref{appendix1:expansion} are not satisfied, the applicability of the LZ formula fails, as discussed in section~\ref{sec:5}.

The uniform asymptotic expansion when both the parameter $|\eta|$ and exponent $|\xi|$ are large was studied thoroughly by Olver \cite{Olver}. By doing $\eta = \frac{1}{2}(\mu^2-1)$ and $\xi = \sqrt{2}\mu \zeta$, provided that $ \arg \mu \in (0, \pi/2)$, then
\begin{eqnarray}\label{appendixA:Asymptotic_olver_1}
	\fl		D_{\frac{1}{2}(\mu^2-1)}(\sqrt{2}\mu \zeta)  = &\frac{2g(\mu)}{(1-\zeta^2)^{1/4}}
	\Big[\cos (\mu^2 \rho - \pi/4)	\sum_{s = 0}^{\infty} \frac{(-1)^s \mathscr{U}_{2s}(\zeta)}{(1-\zeta^2)^{3s}\mu^{4s}} \nonumber \\
	& \qquad \qquad -  \sin (\mu^2 \rho - \pi/4)\sum_{s = 0}^{\infty} \frac{(-1)^s\mathscr{U}_{2s+1}(\zeta)}{(1-\zeta^2)^{3s + 3/2}\mu^{4s+2}}\Big].
\end{eqnarray}
The terms used in \eqref{appendixA:Asymptotic_olver_1} are listed below:
\begin{eqnarray}\label{appendixA:Asymptotic_olver_2}
	g(\mu)&  =  2^{-\frac{1}{4}(\mu^2\!  + 1)}\ee^{-\frac{\mu^2}{4}} \mu^{\frac{1}{2}(\mu^2\! -1)} \Big(1\!-\!\frac{1}{24\mu^2}\! +\! \frac{1}{576 \mu^4}\! +\! \ldots\Big)\\ 
	\rho& =  \frac{1}{2}\Big(\arccos\zeta- \zeta \sqrt{1-\zeta^2}\Big)\\
	\mathscr{U}_{s}& = \mathscr{A}_{s}(\zeta^2-1)^{\frac{3}{2}s}, s \geq 0\\
	\mathscr{A}_0& = 1 \\
	\mathscr{A}_1& = \frac{\zeta^3 - 6\zeta}{24(\zeta^2-1)^{3/2}} \\
	\mathscr{A}_2& =\frac{-9\zeta^4 +249\zeta^2 + 145}{1152(\zeta^2-1)^{3}}  \\
	\mathscr{A}_{s+1}& = \frac{1}{2\sqrt{\zeta^2-1}}\frac{d\mathscr{A}_s}{d\zeta}\! +\! \frac{1}{8}\! \int \!
	\frac{3\zeta^2 +2}{(\zeta^2-1)^{5/2}} \mathscr{A}_s d \zeta,\ s\geq 2.
\end{eqnarray}

\section{The hypergeometric function}
\label{app:B}

Let us recall that the hypergeometric function ${}_2F_1\big(a,b;c;z\big)$ of variable $z$ and parameters $a,b,c,$ is a solution of Euler's hypergeometric differential equation \cite{Abram}
\begin{equation}\label{appendix2:Eulers_hyperg_eq}
	z(1-z) \frac{d^2 w}{dz^2} + \big[c-(a+b+1)z\big]\frac{d w}{dz} - ab z = 0,
\end{equation}
which has three regular singular points: $z =  0,1$ and $\infty$. A particular solution of \eref{appendix2:Eulers_hyperg_eq} around $z = 1$ provided that $c-a-b \notin \mathbb{Z}$ is given by the linear combination of functions
\begin{eqnarray}
	w_1(z) &= {}_2F_1\big(a,b;a+b-c+1; 1-z \big) \\
	w_2(z) &= (1-z)^{c-a-b}{}_2F_1\big(c-a,c-b;1-a-b+c; 1-z\big),
\end{eqnarray}
whose Wronskian \cite{Andr} reads as
\begin{equation}\label{appendix2:wronskian}
	W_{z}\big\{w_{1}(z),w_{2}(z)\big\} = (a+b-c)(1-z)^{-(a+b-c+1)}z^{-c}.
\end{equation}

In Section \ref{sec:caseB} we have functions of the kind ${}_2F_1\big(\alpha + \lambda,\alpha + \lambda;\gamma;x\big)$, which we would like to express in terms of a series when $|\lambda|$ is large. 

When the modulus of parameters $a$ and $b$ of ${}_2F_1\big(a,b;c;z\big)$ is large, we can consider the transformation (when possible)
\begin{eqnarray}\label{appendixB:connection}
	\fl\qquad	{}_2F_1\big(a,b;c;z\big) =&
	\frac{\Gamma(c)\Gamma(a\!+\!b\!-\!c)}{\Gamma(a)\Gamma(b)}\big(1-z\big)^{c-a-b}  {}_2F_1\big(c\!-\!a,c\!-\!b;1\!+\!c\!-\!a\!-\!b;1-z\big) \nonumber \\
	& +  \frac{\Gamma(c)\Gamma(c\!-\!a\!-\!b)}{\Gamma(c\!-\!a)\Gamma(c\!-\!b)}\,{}_2F_1\big(a,b;a\!+\!b\!+\!1\!-\!c;1-z\big)
\end{eqnarray}
where now all three parameters of ${}_2F_1$ are large in modulus. An asymptotic expansion for this case was extensively studied by Watson \cite{Watson}. However, that expansion works in general when the argument is complex, or real but in a certain range. Otherwise the used conventions are not uniquely defined. In section \ref{sec:5} the functions \eref{case2_Q0} or \eref{case2_Q1}, which we want to expand, have a strictly positive argument. Therefore Watson's expansion cannot be applied. Although, Watson's expansion can be extended for real and positive argument, which, along with \eref{appendixB:connection} yields to the following asymptotic expansion
\begin{eqnarray} \label{app2:Expansion_caseB}
	\fl	{}_2F_1\big(\alpha\!+\!\lambda,\alpha\!+\!\lambda;\gamma;z\big) = \frac{2^\gamma (1\!-\!\ee^{-\chi})^{-\frac{1}{2}}(1\!+\!\ee^{-\chi})^{\frac{1}{2}-\gamma}(1\!-\!z)^{-\alpha-\!\lambda}}{\Gamma(\alpha\!+\!\lambda)\Gamma(\gamma\!-\!\alpha\!-\!\lambda)\sin\!\big(\pi(2\alpha\!+\!2\lambda\!-\!\gamma)\big)} \sum_{k=0}^{\infty} \lambda^{-k-\frac{1}{2}}\Gamma\Big(k\!+\!\frac{1}{2}\Big) \nonumber\\
	\fl\qquad\, \qquad\,\qquad\,\qquad\, \,\Big[ \sin\big(\pi(\alpha\!+\!\lambda\!-\!\gamma)\big)C^{(1)}_{k} 
	+\ii(-1)^{k+1}\sin\big(\pi(\alpha\!+\!\lambda)\big)C^{(2)}_{k}\Big],
\end{eqnarray}
valid when $0<z<1$, $|\arg \lambda|  < \pi$ provided that $2(\alpha+\lambda) -\gamma \notin\mathbb{Z}$ and $|\lambda|$ is large. The parameters are defined below:
\begin{eqnarray}
	\chi &= {\rm{arccosh}}(x), \quad \ee^{-\chi} = -|x|+\sqrt{|x|^2-1}\\
	x &= 1-2/(1-z)\\
	C^{(1)}_{k} &= \ee^{-\lambda{\rm{Re}}\{\chi\}-\alpha(i\pi + \xi)}c^{(1)}_{k}, \quad 	C^{(2)}_{k} = \ee^{\lambda{\rm{Re}}\{\chi\}-(\gamma-\alpha)(i\pi + \xi)}c^{(2)}_{k} \\
	c^{(1)}_{0} &= c^{(2)}_{0} = 1\\
	c^{(i)}_{1} &= \frac{1}{2} \frac{L^{(i)}+M^{(i)}\ee^{-\xi} + N^{(i)}\ee^{-2\xi}}{1-\ee^{-2\xi}}, \quad i = 1, 2\\
	L^{(i)} &= (\gamma-1)^2 +(-1)^i(2\alpha-\gamma) - 1/2, \quad i = 1, 2\\
	M^{(1)} &= M^{(2)} = -2(\gamma -1)^2\\
	N^{(i)} &= (\gamma-1)^2 +(-1)^i (\gamma-2\alpha) + 1/2, \quad i = 1, 2.
\end{eqnarray}
Other values of $c^{(i)}_{k}$ for $k \geq 2$ can be derived from the previous expansion and connection formula \eref{appendixB:connection} following \cite{Watson}.

\section{Adiabatic perturbation theory}
\label{app:C}

The APT developed in reference~\cite{Ortiz} is a rigorous method to solve the time-dependent Schr{\" o}dinger equation with a general driven Hamiltonian $\hat{H}\bigl(\boldsymbol{\Lambda}(t)\bigr)$, where $\boldsymbol{\Lambda}(t)$ is a set of parameters varied with a prescribed time dependence, for large total driving time~$T$. The solution is searched as an expansion in powers of $T^{-1}$,
\begin{equation}\label{series}
	\ket{\psi(t)} =\lim_{P\to\infty} {\cal N}_P(t)\sum_{p=0}^{P}T^{-p}\ket{\psi^{(p)}(t)},
\end{equation}
where ${\cal N}_P(t)$ is a normalization coefficient for the series with the maximal order~$P$. The $p$th-order contribution to the state vector, expressed in the instantaneous eigenbasis $\ket{E_n(\boldsymbol{\Lambda}(t))}$ ($n=0,1,\dots,d\!-\!1$) of $\hat{H}\bigl(\boldsymbol{\Lambda}(t)\bigr)$, is given by
\begin{equation}\label{psip}
	\ket{\psi^{(p)}(t)} \equiv \sum_{n = 0}^{d-1} \ee^{-\ii\phi_{n}(t)}\ b_{n}^{(p)}(t)\ \ket{E_n(\boldsymbol{\Lambda}(t))},
\end{equation}
with ${\phi_n(t)=\omega_{n}(t)-\gamma_{n}(\boldsymbol{\Lambda}(t))}$ standing for the total phase composed of the dynamical phase $\omega_n$ and the geometrical phase $\gamma_n$, and $b_{n}^{(p)}(t)$ denoting the coefficients (transition amplitudes) to be determined.
For the initial state $\ket{\psi(0)}=\ket{E_0(\boldsymbol{\Lambda}(0))}$ the lowest-order amplitude reads $b_{n}^{(0)}(t)=\delta_{n0}$, so that the ${p=0}$ term of \eqref{psip} coincides with the adiabatic approximation. In \cite{Ortiz}, a general recursive procedure for finding the amplitudes $b_{n}^{(p)}(t)$ with ${p>0}$ was developed and explicit formulae for low-$p$ cases presented. In \cite{We} it was shown that in this framework the final infidelity is given by
\begin{equation}
	\label{aptinfi}
	{\cal I}^{\rm APT}(T)=\frac{{\cal I}^{(2)}(T)}{T^{2}}+\frac{{\cal I}^{(3)}(T)}{T^{3}}+\frac{{\cal I}^{(4)}(T)}{T^{4}}+\dots,
\end{equation}
while some coefficients ${\cal I}^{(p)}(T)$ with low $p$ were explicitly expressed. For our two-level ($d=2$) system with $\boldsymbol{\Lambda}\equiv\boldsymbol{r}$, the leading order term in \eqref{aptinfi} is
\begin{eqnarray} 
	\label{APT:two_level}
	\fl\quad 
	\frac{{\cal I}^{(2)}(T)}{T^2}=\frac{v^2(T)}{\Delta E^2(\boldsymbol{r}(T))}+\frac{v^2(0)}{\Delta E^2(\boldsymbol{r}(0)))}   
	-2\sum_{\mu,\nu} \dot{\mu}(T) \dot{\nu}(0)\times
	\\
	\fl
	\times{\rm Re}\biggl[\ee^{-\ii[\phi_1(T)-\phi_0(T)]}\frac{\matr{E_{0}(\boldsymbol{r}(T))}{\frac{\partial\hat{H}}{\partial\mu}(\vecb{r}(T))}{E_{1}(\boldsymbol{r}(T))}}{\Delta E^2(\boldsymbol{r}(T))}\frac{\matr{E_{1}(\boldsymbol{r}(0))}{\frac{\partial\hat{H}}{\partial\nu}(\vecb{r}(0))}{E_{0}(\boldsymbol{r}(0))}}{\Delta E^2(\boldsymbol{r}(0))}\biggr],
	\nonumber
\end{eqnarray}
where $v(t)$ is the metric speed \eqref{speed_on_manifold} and $\mu,\nu\in\{x,y,z\}$ or $\{r,\vartheta,\varphi\}$.

\end{document}